\documentclass[letterpaper,journal]{IEEEtran}
\usepackage{cite}
\usepackage{xcolor}
\usepackage[colorlinks=false, pdfborder={0 0 0}]{hyperref}
\usepackage{multirow}
\usepackage{algorithm,algorithmic}
\usepackage{amsmath,amssymb,amsfonts}
\usepackage{algorithmic}
\usepackage{array}
\usepackage[caption=false,font=normalsize,labelfont=sf,textfont=sf]{subfig}
\usepackage{textcomp}
\usepackage{stfloats}
\usepackage{url}
\usepackage{verbatim}
\usepackage{hyperref}
\usepackage{graphicx,color}
\usepackage{xurl}      
\usepackage{hyperref}
\def\BibTeX{{\rm B\kern-.05em{\sc i\kern-.025em b}\kern-.08em
    T\kern-.1667em\lower.7ex\hbox{E}\kern-.125emX}}
\usepackage{balance}

\begin{document}

\title{Machine Learning-based Path Loss Prediction in Suburban Environment in the Sub-6 GHz Band}

\author{Ferdaous Tarhouni, Muneer Al-Zubi, Member, IEEE, and Mohamed-Slim Alouini, Fellow, IEEE

\thanks{The authors are with the Computer, Electrical and Mathematical Science and Engineering Division (CEMSE), King Abdullah University of Science and Technology (KAUST), Thuwal 23955-6900, Makkah Province, Saudi Arabia (e-mail: ferdaous.tarhouni@kaust.edu.sa; muneer.zubi@kaust.edu.sa; slim.alouini@kaust.edu.sa).}}

\markboth{}%
{How to Use the IEEEtran \LaTeX \ Templates}

\maketitle

\begin{abstract}
Accurate path loss (PL) prediction is crucial for successful network planning, antenna design, and performance optimization in wireless communication systems. Several conventional approaches for PL prediction have been adopted, but they have been demonstrated to lack flexibility and accuracy. In this work, we investigate the effectiveness of Machine Learning (ML) models in predicting PL, particularly for the sub-6 GHz band in a suburban campus of King Abdullah University of Science and Technology (KAUST). For training purposes, we generate synthetic datasets using the ray-tracing simulation technique. The feasibility and accuracy of the ML-based PL models are verified and validated using both synthetic and measurement datasets. The random forest regression (RFR) and the K-nearest neighbors (KNN) algorithms provide the best PL prediction accuracy compared to other ML models. In addition, we compare the performance of the developed ML-based PL models with the traditional propagation models including COST-231 Hata, Longley-Rice, and Close-in models. The results show the superiority of the ML-based PL models compared to conventional models. Therefore, the ML approach using the ray-tracing technique can provide a promising and cost-effective solution for predicting and modeling radio wave propagation in various scenarios in a flexible manner.
\end{abstract}


\vspace{-1mm}

\section{Introduction}
Understanding and characterizing radio wave propagation is crucial for planning and optimizing mobile communication networks, particularly in the frequency bands of 5G and 6G networks \cite{1}. The radio signal experiences various impairments in the propagation channel due to physical mechanisms, e.g., reflection, diffraction, and scattering, and other environmental factors, e.g., rain and gases, that impact signal propagation in mobile communication systems. This will cause an attenuation (i.e., power reduction) in the radio signal at the mobile receiver which is known as path loss \cite{2}. Therefore, precise prediction of path loss (PL) is necessary for link budget analysis and optimizing the system design parameters that can reduce the cost and time of mobile network deployments \cite{3,4}.

Several radio frequency (RF) propagation models have been assessed in various frequency bands for different conditions and environments. The conventional models can be categorized as deterministic, empirical, and stochastic modelling approaches. Each method has its own set of benefits and limitations regarding computational efficiency, applicability, and accuracy. Empirical methods are fundamentally derived based on measurements conducted at certain frequencies and locations such as log-distance, Egli, Hata, and Okumura models \cite{5,6}. In \cite{7,8}, the performance of various empirical PL models including the Okumura-Hata, the COST-231 Hata, the ECC, and the Ericsson models, are examined in urban, suburban, and rural areas. In \cite{8,9}, the Lee model was tuned for cellular communications at different locations using the least squares (LS) method. In \cite{10}, the authors optimized different empirical PL models based on extensive radio measurements conducted at different frequency bands of the global system for mobile communications (GSM) in urban areas in Irbid city, Jordan. These models use simple mathematical equations that are less precise when used in other locations and situations since they are derived from observations made at a specific site.  

The deterministic models, derived based on electromagnetic theory and theoretical physics, provide more accurate PL prediction for each specific site using the terrain and clutter data as main input parameters. The most used deterministic modelling approach is the ray-tracing technique \cite{12}. However, the ray-tracing model requires intensive computational time and resources, particularly, for long propagation paths. In \cite{13}, a radio propagation model for macrocell coverage forecasts at the sub-6 GHz frequency band is presented using a hybrid modelling approach. The hybrid model is proposed based on the building-transmission model (BTM), the international telecommunication union of radio-communications (ITU-R) model, and the discrete mixed Fourier transform split-step parabolic equation (DMFT-SSPE). The accuracy of the suggested hybrid approach is assessed in comparison to real data taken from a comprehensive measurement campaign carried out in Rio de Janeiro, Brazil. 

Stochastic models, on the other hand, represent the environment using random variables. While they are generally less accurate than deterministic models, they require minimal environmental information and significantly lower computational resources to generate predictions \cite{abhayawardhana2005comparison}. A comprehensive review of PL models for indoor millimeter-wave propagation has been proposed in \cite{oladimeji2022propagation}. The study compares the close-in (CI), floating intercept (FI), and Alpha-Beta-Gamma (ABG) models. It emphasizes the need for improved methods to address the high PL and variability in indoor 5G environments, proving that CI and FI offer better accuracy and simplicity for both line-of-sight (LoS) and non-line-of-sight (NLoS) scenarios. In \cite{elmezughi2021efficient}, enhanced versions of the CI and FI models have been explored by incorporating an additional term to improve prediction accuracy without increasing model complexity. Using real measurements at 14, 18, and 22 GHz, the study shows improved fitting to indoor corridor data, achieving reduced shadow fading and better stability under distinct system parameters such as antenna height and angle of arrival. In \cite{sun2016investigation}, a comparison has been provided between ABG, CI, and CI with a frequency-weighted PL exponent (CIF) across 30 datasets with frequencies ranging from 2 to 73 GHz in diverse environments. It has been demonstrated that CI and CIF models ensure better parameter stability, lower prediction errors, and simpler implementation than ABG, particularly in outdoor settings.

Although some traditional RF propagation models, such as empirical and stochastic models,  offer advantages in terms of simplicity, they provide less accurate results and often lack generalizability and adaptability when applied to varying scenarios. Moreover, although deterministic models (e.g., ray-tracing)  provide accurate predictions, they require extensive computational complexity. These inherent limitations motivate the adoption of machine learning (ML) techniques, which can effectively capture complex environmental patterns and offer flexibility across diverse propagation environments.  Therefore, we can have an accurate ML model with less computational complexity than deterministic models. The computational demand of a machine learning model lies almost entirely in its training; once deployed, it operates with minimal computational time. Indeed, ML-based
models have garnered widespread attention within the research community as promising alternatives to conventional PL prediction models. In \cite{14}, a general overview of recent developments in PL modelling, based on ML approaches, has been discussed. Using deep learning (DL) techniques, mainly convolutional neural networks (CNN) combined with satellite imagery and position indicators, an improved PL prediction model for mobile communication systems at 2.6 GHz is proposed in \cite{16}. In \cite{17}, a modelling approach based on three main techniques: multi-dimensional regression using artificial neural networks (ANN), variance analysis using Gaussian processes, and feature selection aided by principal component analysis (PCA), has been proposed to model PL at three frequencies 450, 1450, and 2300 MHz in the suburbs of a small town called Nonsan. In \cite{18}, the authors applied a model-aided DL approach to estimate the PL for the 6 GHz band using data collected in four locations in Boulder and Louisville, in Colorado, under LoS/NLoS scenarios. Multi-layer perceptron (MLP) neural network is proposed to predict PL and analyze the effect of the environmental features on the prediction process \cite{19}. The evaluation has been performed using measurements taken at 2.5 GHz in three different locations in Hangzhou, China.

In \cite{9968107}, the authors proposed an ML-based approach using CNNs for PL prediction in urban environments in Munich, Paris, and Singapore. By leveraging available geographic data from OpenStreetMap and other geographic information system (GIS) sources, this method significantly outperforms traditional ray-tracing methods in terms of prediction speed while maintaining good accuracy across frequencies ranging from 897 MHz to 60 GHz. On the other hand, a hybrid method leveraging both ML models and ray-tracing tool has been explored in \cite{15}. In fact, the PL has been evaluated in urban environments in Frankfurt for both LoS and NLoS scenarios using various ML techniques,  including support vector regression (SVR), random forest regression (RFR), and K-nearest neighbor (KNN) algorithms. The training and testing phases were conducted using a dataset created by ray-tracing simulations of a long-term evolution (LTE) network operating at 2.1 GHz. The COST231 Walfisch-Ikegami empirical model has been used for comparison reasons in order to validate the high performance of the proposed method.

Although the aforementioned studies acknowledge the efficiency of ML models in PL prediction, they still exhibit limitations. Many of these works depend primarily on experimental measurements, which are generally costly, time-consuming, and challenging to carry out. Additionally, authors often concentrate on predictions at a single frequency or a limited set of frequencies, limiting the broader applicability of the work. Furthermore, according to existing literature, only a few studies have explored hybrid approaches that integrate both ML methods and ray-tracing simulations, indicating a gap that this work aims to fill.
\par In this paper, we exploit the capability of ML to accurately predict the PL in a suburban area for 5G mid-band spectrum (1-6 GHz) based on a synthetic PL dataset, generated by ray-tracing technique, and perform validation using outdoor measurements. In this work, ray-tracing tool is used to generate the training and testing datasets used for the ML-based approach for different sites inside the campus of the King Abdullah University of Science and Technology (KAUST). The KAUST campus can be considered a suburban area consisting of both university buildings and residential houses of different heights and sizes.  The datasets have been generated for three transmitter sites with various values of the transmitter height, transmitter power, frequency, etc.  The ML models are trained and tested over various frequencies used in the mobile networks including 1.5, 2.3, 2.5, 3.5, and 6 GHz. Then, the ML-based PL model is verified and validated with experimental measurements carried out in Rio de Janeiro city, Brazil \cite{13}. In addition, a comparative analysis is performed between the ML-based PL model and some conventional PL models to highlight the performance and feasibility of the ML algorithms compared to the traditional approaches.
\par The remainder of this paper is organized as follows. In Section \ref{2}, we describe the modeling framework including the modeling techniques, i.e., conventional and ML-based approaches. In Section \ref{3}, we illustrate the data generating process. In Section \ref{4}, we show and discuss the obtained results using both ML techniques and traditional models, compared with measured data. Finally, the paper is concluded in Section \ref{5}.
\section{Modeling Framework}
\label{2}
\par In this section, we present the overall modeling approach, which includes an overview of both the traditional RF propagation models used for comparison and the ML-based algorithms employed to build an optimized PL prediction model. We begin by introducing the conventional models, widely applied for PL prediction but, as previously mentioned, often unable to account for the complexity and variability of real-world environments. These models serve as baselines for performance evaluation. To assess the effectiveness of the ML-based approach, which will be explored in the following section, we compare its performance against the traditional methods adopting several evaluation metrics. This comparison quantifies the prediction error and demonstrates the accuracy of the proposed ML models.
\vspace{-3mm}

\subsection{Conventional RF Propagation Models}
This subsection provides an overview of the traditional PL models used for comparative purposes with the ML-based models, including the ray-tracing, Longley-Rice, Close-in, and COST-231 Hata models.

\vspace{1mm}

\subsubsection{Ray-Tracing Propagation Model}
Radio propagation modeling based on the ray-tracing technique is classified as a deterministic modeling approach \cite{20}. It is worth noting that the accuracy of ray-tracing primarily depends on the quality of input parameters, particularly high-resolution digital terrain and clutter data, which allow precise prediction of path loss. With the widespread availability of such data online, ray-tracing can be applied reliably across diverse regions worldwide. Unlike empirical models, ray-tracing models are specific to a given 3-D environment and thus suitable for urban and suburban scenarios. Regarding the MATLAB ray-tracing tool, it supports both indoor and outdoor 3D environments and operates across a broad frequency range from 100 MHz to 100 GHz, making it highly versatile for planning and designing wireless communication systems. In the ray-tracing method, the radio wave is approximated as a large number of very narrow rays that carry energy. These rays interact with the environment in several ways including the direct path (i.e., LoS), refraction, reflection, diffraction, and scattering. Ray-tracing models utilize numerical simulations to predict the trajectories of rays from transmitter to receiver in a 3-D environment. The model derives crucial parameters from these paths including the angle of departure/arrival and the time of arrival (delay) and correctly models multipath propagation, including variations in signal strength around corners and behind buildings or hills. Then, the total PL is estimated as a sum of interaction losses, free space loss, and atmospheric loss. There are two ray-tracing methods, i.e., the image method and the shooting and bouncing rays (SBR) method, that vary in the types of interactions, computational speed, and accuracy. The SBR method is generally less accurate compared to the image method, but it is faster. Moreover, the SBR method includes effects from reflection and diffraction while the image method can only model the reflection. Therefore, we use the SBR method in this work to generate the training and testing PL datasets. 

\vspace{3mm}
\subsubsection{Longley-Rice Model}
The Longley-Rice model, aka, the Irregular Terrain Model (ITM), is a semi-deterministic propagation model that was implemented using the electromagnetic theory and radio measurements \cite{21}. It is used to predict radio wave propagation over irregular terrain considering parameters including carrier frequency, atmospheric conditions, distance, antenna heights, and terrain elevation. This model is suitable for frequencies between 20 MHz and 20 GHz, antenna heights between 0.5 m and 3000 m, and distances between 1 km to 2000 km. In addition to free space loss, this model calculates the PL due to various physical phenomena including reflection, scattering, and diffraction. This model yields critical information like received signal power, coverage areas, and other propagation characteristics, offering valuable insights for planning and deploying wireless communication networks. The Longley-Rice model is used in various radio propagation software such as Radio Mobile \cite{radiom}, SoftWright TAP$^{\textnormal{TM}}$ \cite{softw}, and communication toolbox in MATLAB. In this paper, we use the Longley-Rice model in the MATLAB communication toolbox to predict the PL.
\vspace{3mm}
\subsubsection{Close-in Path Loss Model}
The Close-in (CI) PL model is a multi-frequency statistical (i.e., stochastic) model that characterizes large-scale propagation PL in a given environment \cite{sun2016investigation}. This model includes a free-space term as a function of frequency and a standard 1-m free-space reference distance. This term leads to higher parameter stability and better prediction accuracy when utilizing the model outside of the range of its measurements. The estimated PL using the CI model is given by \cite{22}:
\begin{equation}
\begin{aligned}
\mathrm{PL}^{\mathrm{CI}}(f,d)\,[\mathrm{dB}]
&= \mathrm{FSPL}(f,d_0)\,[\mathrm{dB}]
  + 10\,n \log_{10}\!\left(\frac{d}{d_0}\right) \\
&\quad + \chi_\sigma^{\mathrm{CI}} \, .
\end{aligned}
\end{equation}

where $d \geq d_0$, $f$ is the frequency in GHz, $d_0$ is the close-in free-space reference distance in m, $n$ is the PL exponent (PLE), and $\chi_\sigma^{\mathrm{CI}}$ is a zero-mean Gaussian random variable with a standard deviation $\sigma$ in dB, $d$ is the 3-D separation distance between the transmitter and receiver in m, and $FSPL(f,d_0)$ is the free-space path loss (FSPL) in dB at $d_0$ and $f$ given as \cite{22}:
\begin{equation}
\operatorname{FSPL}\left(f, d_0\right)[\mathrm{dB}]=20 \log _{10}\left(\frac{4 \pi f d_0 \times 10^9}{c}\right),
\end{equation}
\noindent where $c$ is the light speed.

In this paper, we use the CI model in the MATLAB communication toolbox to predict the PL for the purpose of comparison with other models.

\vspace{3mm}

\subsubsection{COST-231 Hata Model}
The COST-231 Hata model is an empirical PL model implemented as an extension version of the Okumura-Hata model to extend the maximum applicable frequency range up to 2000 MHz \cite{23}. It is widely used for PL prediction in cellular mobile systems for various types of terrains including rural, suburban, and urban areas. There are various radio propagation software that use this model with some modifications. This model is applicable for the following range of parameters: frequency of 1500-2000 MHz, distance of 1-20 km, BS antenna height of 30-200 m, and mobile antenna height of 1-10 m. Although this model is valid for BS antenna heights greater than 30 m, still it can be used for lower heights assuming that the BS antenna height is higher than the surrounding buildings \cite{24}. 
\par In our work, we assume that the BS antenna is placed at the building rooftop, i.e., the antenna height is higher than the adjacent buildings. Also, a distance range lower than 1km can be allowed in this model with correction factors and optimization process \cite{25,26}. The KAUST campus can be considered as a suburban area with few large buildings and residential areas of many small houses (i.e., less than two floors). The COST-231 Hata model for suburban areas is given as \cite{25}:
\begin{equation}
\begin{split}
 PL_{dB} = &\ 46.3+33.9\log_{10}(f)-13.82\log_{10}(h_{T})-a(h_{R}) \\
 &\ +(44.9-6.55\log_{10}(h_{T}))\log_{10}(d)+C,
 \label{costf}
\end{split}
\end{equation}
\begin{equation}
  a(h_{R})= (1.1\log_{10}(f)-0.7)h_{R}-(1.56\log_{10}(f)-0.8),  
\end{equation}
\begin{equation}
C= 0 \text{ dB},
\end{equation}
where $f$ is the operating frequency in MHz, $h_T$ is the transmitter height in m, $h_R$ is the receiver height in m, $d$ is the separation distance in km.

\vspace{-3mm}

\subsection{Machine Learning Algorithms}

In this subsection, we describe the widely used ML algorithms including ANN, RNN-LSTM, DTR, RFR, and KNN. In these models, eight input variables (features) have been fed from the preprocessed data which consists of the transmitter height, transmitter power, frequency, distance, elevation angle, LOS status, latitude difference, and longitude difference.

\vspace{2mm}

\subsubsection{Artificial Neural Network Model}
\begin{figure}[ht]
\centering
\captionsetup{justification=centering}
\includegraphics[width=96mm]{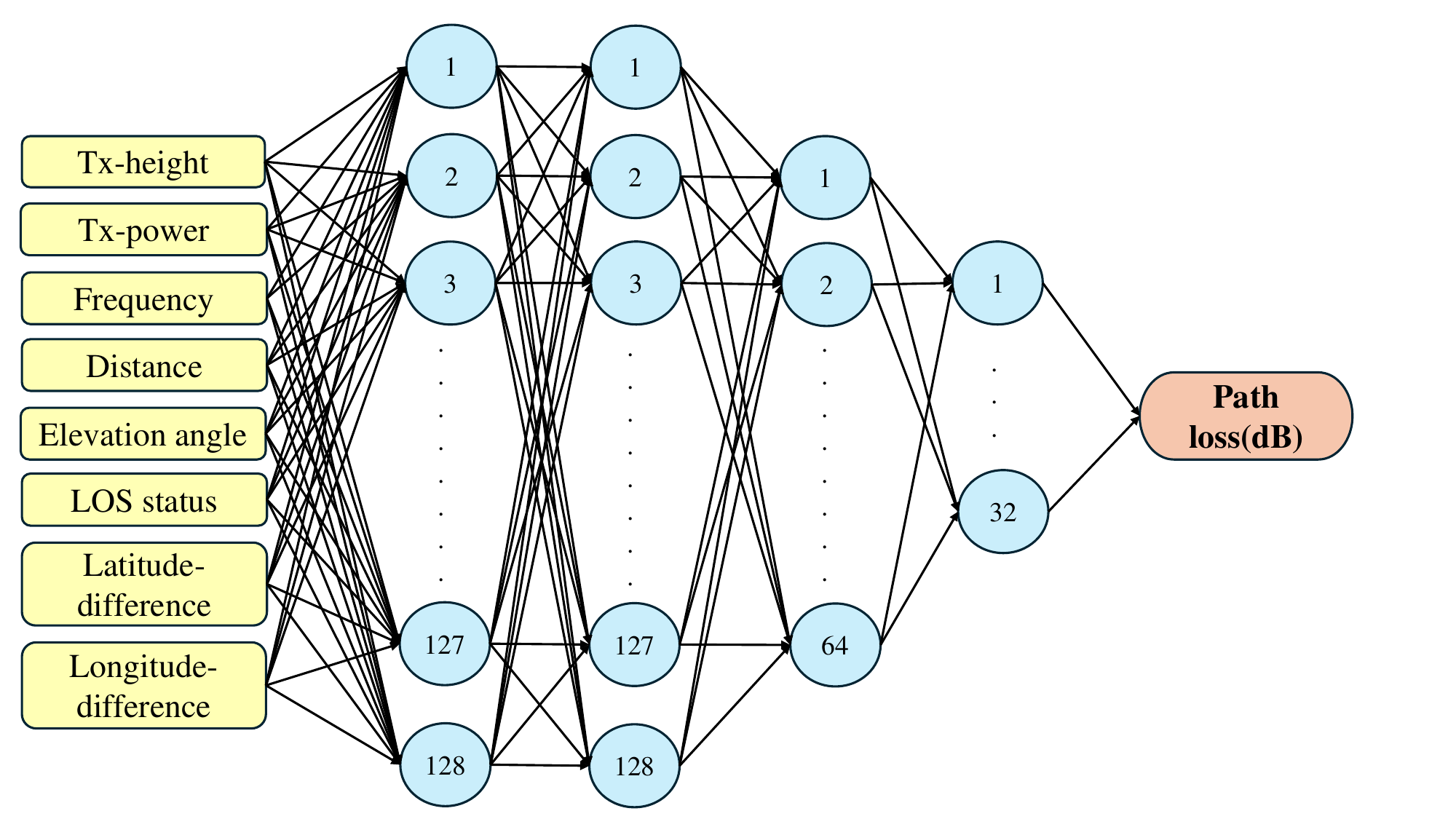}
\vspace{-1em}
\caption{ANN model architecture.}
\label{fig:ANN}
\end{figure}

The artificial neural network (ANN) is a model inspired by the structure of biological neural networks, particularly the human brain \cite{27}. Its architecture comprises interconnected artificial units organized into distinct layers. These layers consist of the input layer, the hidden layers, and the output layer. The ANN model incorporates a fundamental component known as the activation function, which facilitates the learning of complex relationships between features within the network. The hyperparameters that fit the data will be illustrated later. The ANN model architecture is illustrated in Figure \ref{fig:ANN}.
\vspace{3mm}
\subsubsection{Recurrent Neural Network Model}
\begin{figure}[ht]
\centering
\includegraphics[width=95mm]{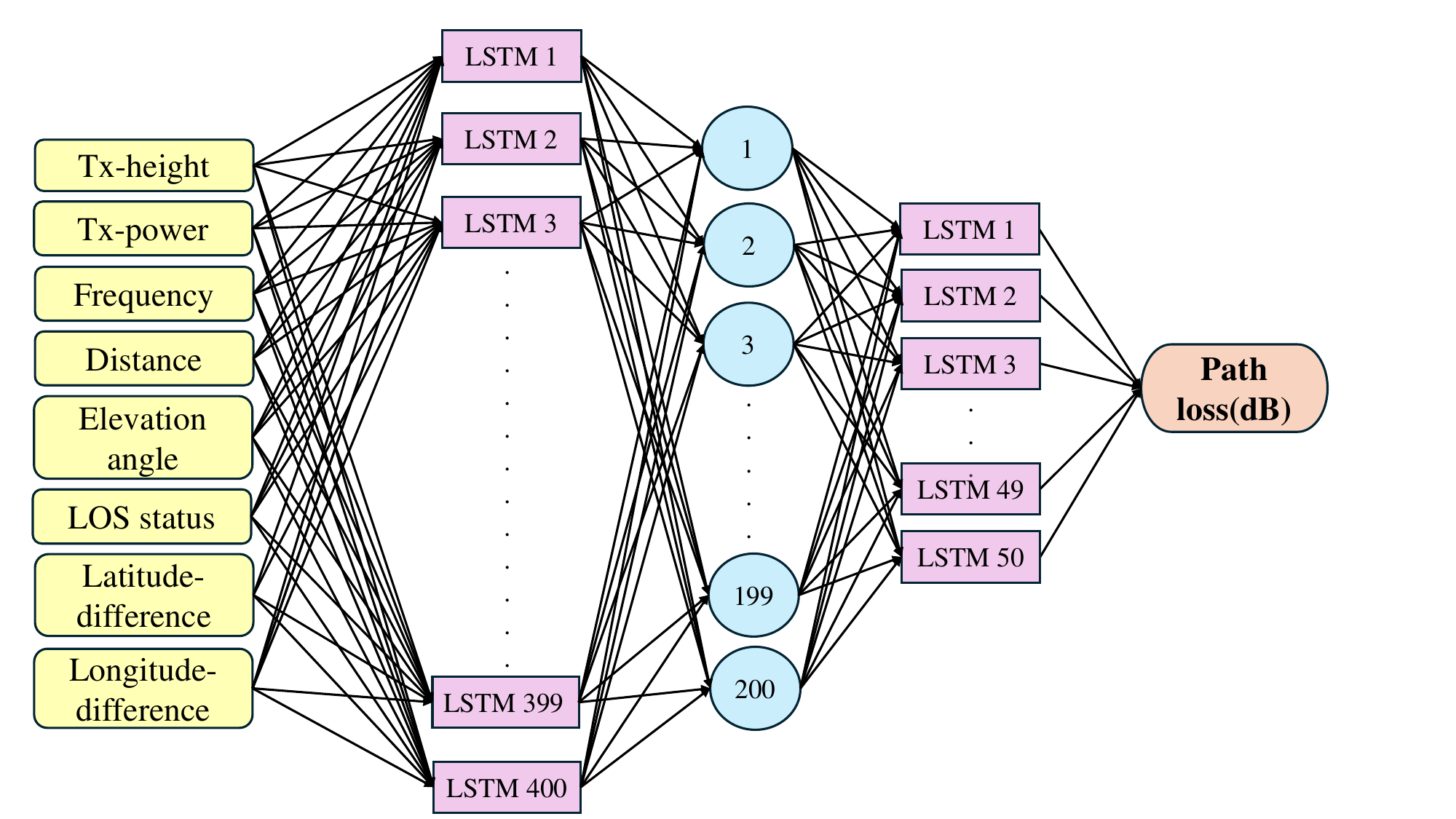}
\vspace{-1em}
\caption{RNN model architecture.}
\label{fig:RNN}
\end{figure}

The recurrent neural network (RNN) is a variation of ANN, where the connections between nodes form a directed graph along a temporal sequence. The RNN is particularly enhanced by the inclusion of the long short-term memory (LSTM) layer. The LSTM introduces a memory cell that possesses the ability to selectively remember and erase information across multiple time steps, thereby enabling the capture of long-term dependencies in sequential data. In practice, the LSTM unit receives input, adjusts its hidden state and memory cell, and produces output data at each time step. The hidden state is then transmitted to the subsequent time step, allowing the network to learn from the entire sequence. The proposed architecture is anticipated to capture feature representations that encode specific aspects of PL. The structure of the proposed model aligns with the original version of the RNN-LSTM, with modifications related to the setup of hyperparameters, as will be presented later. Figure \ref{fig:RNN} shows the architecture of the proposed RNN model.
\vspace{2mm}

\subsubsection{Decision Tree Regression Model}
The decision tree regression (DTR) is a widely used tool in machine learning due to its simplicity and interpretability. This model constructs a binary tree structure, where each leaf node represents a numerical target, and each internal node signifies a decision based on input features. The algorithm seeks to partition the feature space in a way that minimizes the prediction error. During the decision tree construction, the model iteratively selects the best feature and splitting point, minimizing the error according to a specific metric, such as the mean squared error (the default criterion). The splitting process continues until it reaches either the maximum tree depth or the minimum number of samples required to split a node. The key parameter to optimize for this algorithm is the maximum tree depth. Therefore, we searched for the optimal value that yields good results in our study.

\vspace{2mm}

\subsubsection{Random Forest Regression Model}
The random forest regression (RFR) model incorporates multiple random decision trees, each constructed using a distinct split of data which enhances the stability of the model. More specifically, each tree involves a root node, internal nodes, and leaf nodes related to the final results. The trees operate in parallel, and their predictions are subsequently averaged to generate a final output. The main parameters to optimize in our proposed model are the number of estimators and the maximum tree depth. Indeed, the first one sets the ensemble size, while the latter regulates the decision tree’s maximum split number. By averaging the results from each individual decision tree, the predicted value of new samples for PL prediction can be calculated as follows.
\begin{equation}
\label{rf}
\hat{PL}=\frac{1}{T} \sum\limits_{t=1}^T \hat{pl_{t}}(x), 
\end{equation}
where $\hat{pl_{t}}(x)$ is the predicted PL of the $t_{th}$ decision tree model, $x$ is the set of features, $T$ is the number of decision tree models.

\vspace{2mm}

\subsubsection{K-Nearest Neighbors Regression Model}
The K-nearest neighbors (KNN) model is a supervised ML technique used in regression tasks. It is based on a similarity metric when predicting the target. Initially, the model computes the distance separating the input data point and each training point in the data. In terms of distance, the algorithm selects the k closest neighbors to the input data, and then, averages their target values to obtain the predicted value. In this model, both the number of neighbors and the distance are crucial parameters. For instance, the frequently used metrics are the Manhattan distance, the Euclidean distance, and the Minkowski distance. In this work, we use the Manhattan distance.  For the number of neighbors 'K', we searched for the optimal value that leads to the minimum error. Typically, the features have different value ranges and varying effects on distance calculation. In comparison to random forest and decision tree regressors, the KNN algorithm is more sensitive to the inputs and thus it is better to use normalization with this model.

\section{Data Generating Process}
\label{3}
\begin{figure*}[ht]
\centering
\includegraphics[width=150mm]{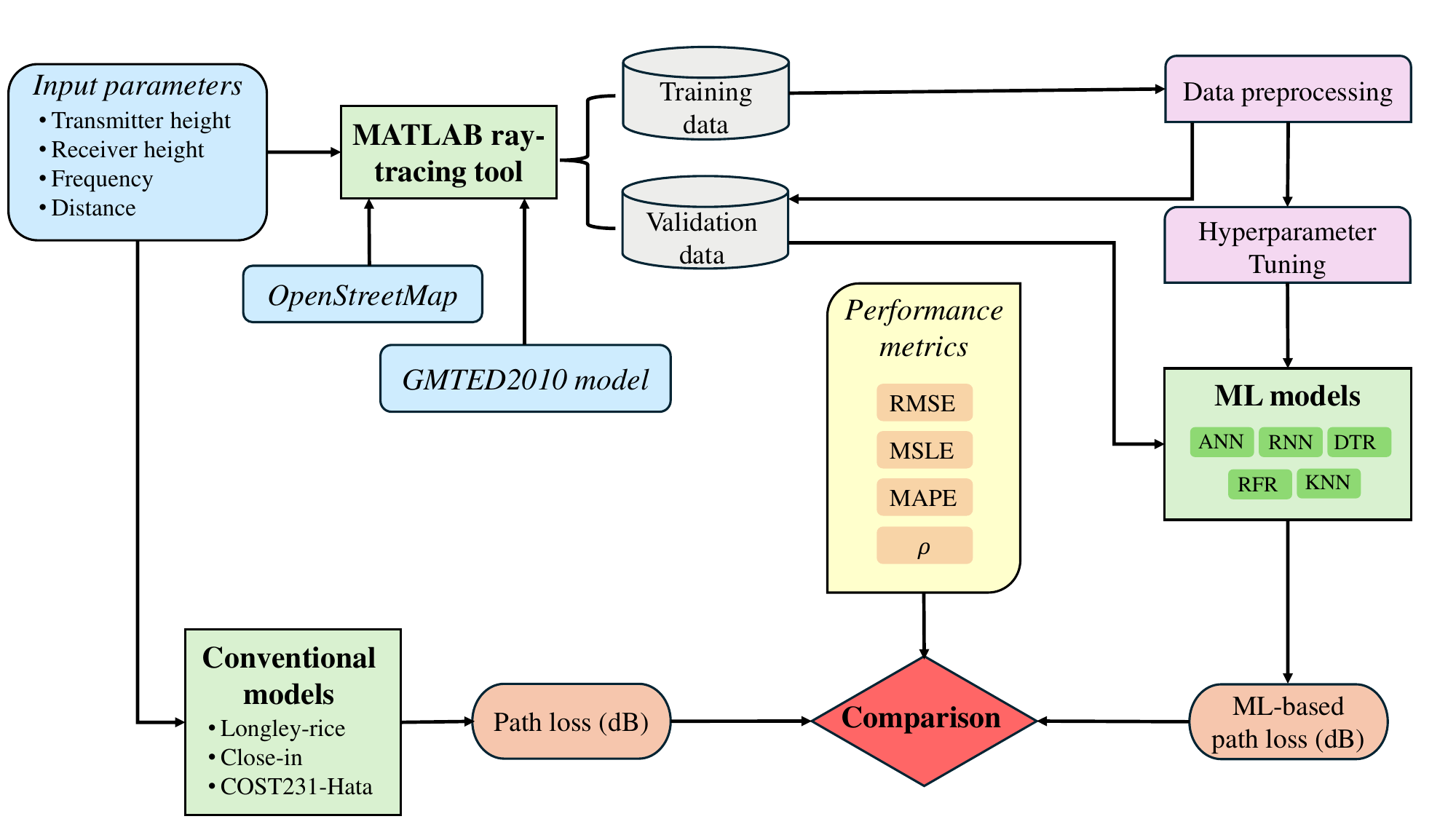}
    \caption{Flowchart of the modeling approach.}
    \label{MLdiagram}
\end{figure*}

\begin{figure}[ht]
\centering
\includegraphics[width=87mm]{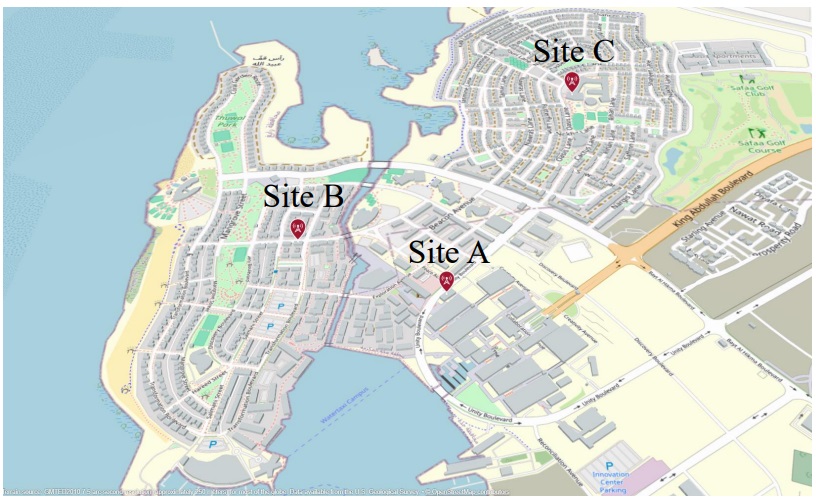}
\caption{The location map of the transmitters under study.}
\label{fig:sites}
\end{figure}

In this section, we demonstrate the overall process used to generate the training and testing PL dataset using the ray-tracing technique, as illustrated in Figure~\ref{MLdiagram}. The region of interest in this work is the KAUST Campus which is a suburban area located at Thuwal in Saudia Arabia that consists of residential areas of many small houses (e.g., villas and townhouses) and few university campus buildings. The area that includes the KAUST campus is extracted using the OpenStreetMap (OSM) geographic database \cite{28}. The extracted map provides approximate information regarding the heights of various structures (i.e., houses, buildings, etc.) based on the number of floors. Then, we chose three distinct locations for the transmitters, i.e., the base station (BSs), as shown in Table \ref{table:sites} and Figure \ref{fig:sites}. The BS antenna is assumed to be placed on a tower of different heights on the building rooftop to extend the coverage area. The extracted open street map is used together with the global multi-resolution terrain elevation data (GMTED2010) model as an input to the MATLAB RF propagation tool. Then, the simulation was performed using the ray-tracing model in MATLAB to obtain the received signal strength in the area surrounding the BSs for various system parameters as listed in Table \ref{table:parameters}. We used different values for the transmitter height, the transmitter power, and the operating frequency. The frequency bands under test were selected within the radio spectrum of mobile networks, i.e., sub-6 GHz band \cite{29}. The receiver height is set at 1.5 m, which is the average height of a human. We adopt a reference transmitter antenna gain of 0 dBi. Since this gain is held constant across all scenarios, it does not affect the final results.
\begin{table}[ht]
\centering
\caption{The geographical location of the transmitters under study.}
\begin{tabular}{|p{0.3\linewidth}| p{0.27\linewidth}| p{0.27\linewidth}|}
\hline
\textbf{Site Name}     & \textbf{Latitude} & \textbf{Longitude}
\\ \hline
Site A (Building 18)  &22.311359  &39.102723\\\hline
Site B (Safaa Gardens)  &22.322863  &39.108194\\ \hline
Site C (Island Street)   &22.313830  &39.097040\\ \hline
\end{tabular}
\label{table:sites}
\end{table}

In this work, we assume that the feeder cable loss between the BS and the antennas is negligible. After obtaining the received signal strength at various locations in the coverage area of each transmitter, the PL is calculated as follows \cite{30}:
\begin{equation}
 P_{L}(d)= P_{Tx} + G_{Tx} + G_{Rx} - P_{Rx}(d),
 \label{p_l}
\end{equation}
where $P_{Tx}$ is the transmitter power in dBm, $P_{L}(d)$ is the PL in dB at distance d, $d$ is the distance between the transmitter and the receiver, $P_{Rx}(d)$ is the receiver power in dBm, $G_{Tx}$ is the transmitter gain in dBi, and  $G_{Rx}$ is the receiver gain in dBi.

The data obtained from the simulation \cite{ferdaous25} include the following parameters at each receiving point inside the coverage area which are used as features for the ML algorithms: the latitude/longitude of the receiving point, the distance between the transmitter and the receiver, the azimuth/elevation angles, the LoS status, the received power, and the PL. These parameters are obtained in the coverage area of three different sites for various values of transmitter height, transmitter power, and frequency, see Table \ref{table:parameters}. Thus, we get a total of 45 different sub-datasets for each site which are then combined into one main dataset for each site with a size of 95,000 samples.  
\begin{table}[ht]
\centering
\caption{System parameters used in the RF propagation models for PL evaluation.}
\begin{tabular}{|p{2cm}|p{1.8cm}|p{0.7cm}|p{2.5cm}|}
\hline
\textbf{Parameter}            & \textbf{Value}  & \textbf{Unit} & \textbf{Description}
\\\hline
\textit{f}            & (1.5, 2.3, 2.5, 3.5, 6) & GHz & Operating frequency  \\ \hline
$h_{Tx}$    & (12, 16, 21) & m & Transmitter height above ground level\\ \hline
$h_{Rx}$ & 1.5 & m & Receiver height\\ \hline
$P_{Tx}$ & (5, 10, 15)  & W & Transmitter power \\ \hline
$N_R$ & 4 & -  & Maximum number of reflections\\ \hline
$d_{max}$ & 1500 & m & Maximum distance range\\ \hline
$G_{Tx}$ & 0 & dBi & Transmitter antenna gain\\ \hline
$G_{Rx}$ & 2.1 & dBi & Receiver antenna gain\\ \hline
Terrain material & concrete & - & -\\ \hline
Building material & concrete & - & -\\ \hline
\end{tabular}
\label{table:parameters}
\end{table}
\par The size of one main dataset is about 95000 samples. The main datasets of sites A and B are used for training while the dataset of site C is used for validation. The datasets of sites A and B are divided into 80\% training data and 20\% testing data. Regarding the deep learning models, we extract 15\% of the training set for validation to tune hyperparameters and mitigate overfitting.

\begin{figure}[ht]
\centering
\begin{minipage}{0.45\textwidth}
\centering
\includegraphics[width=85mm]{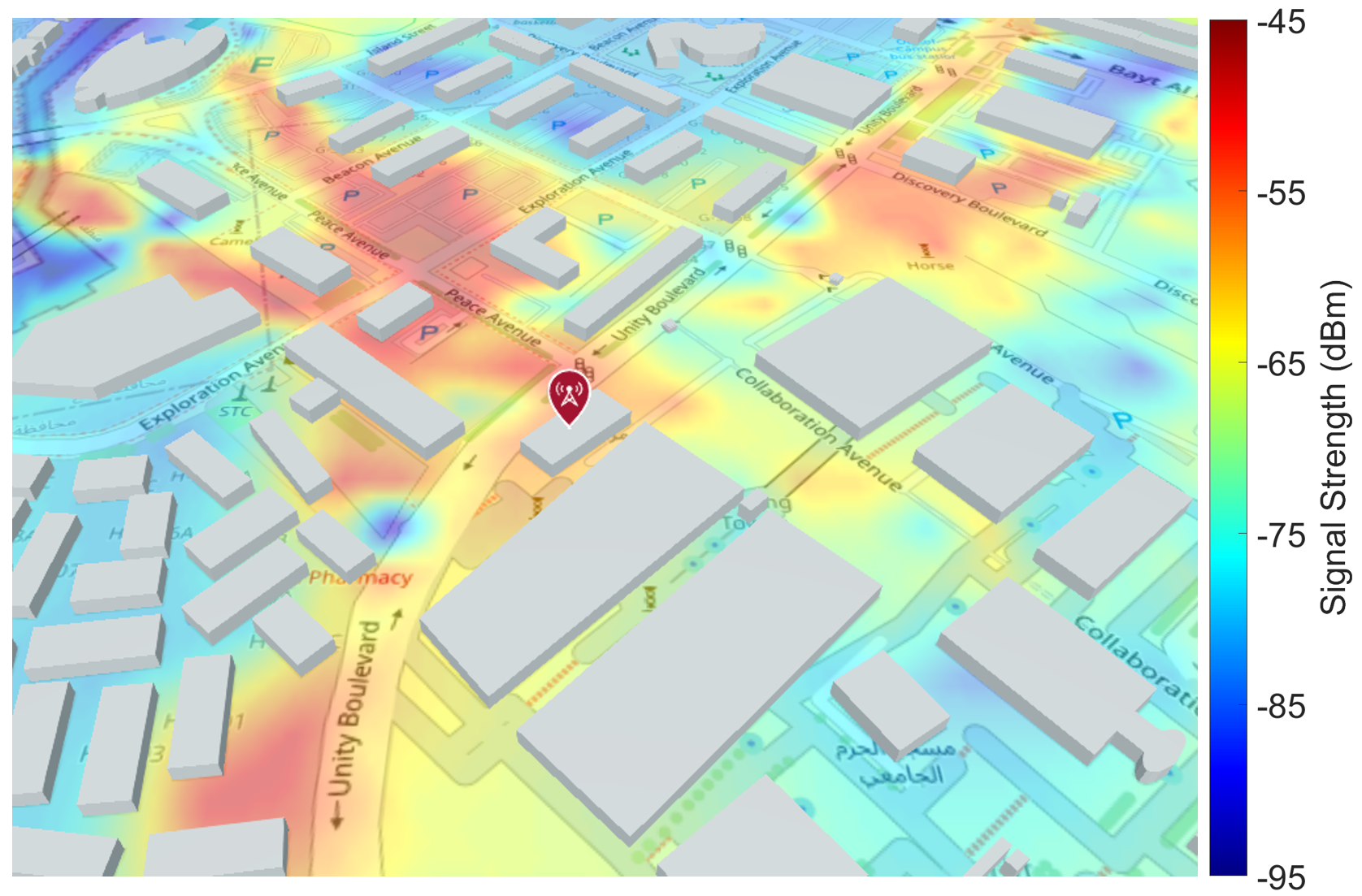}
    \\(a)
\end{minipage}
\hfill
\begin{minipage}{0.45\textwidth}
\centering
\includegraphics[width=86mm]{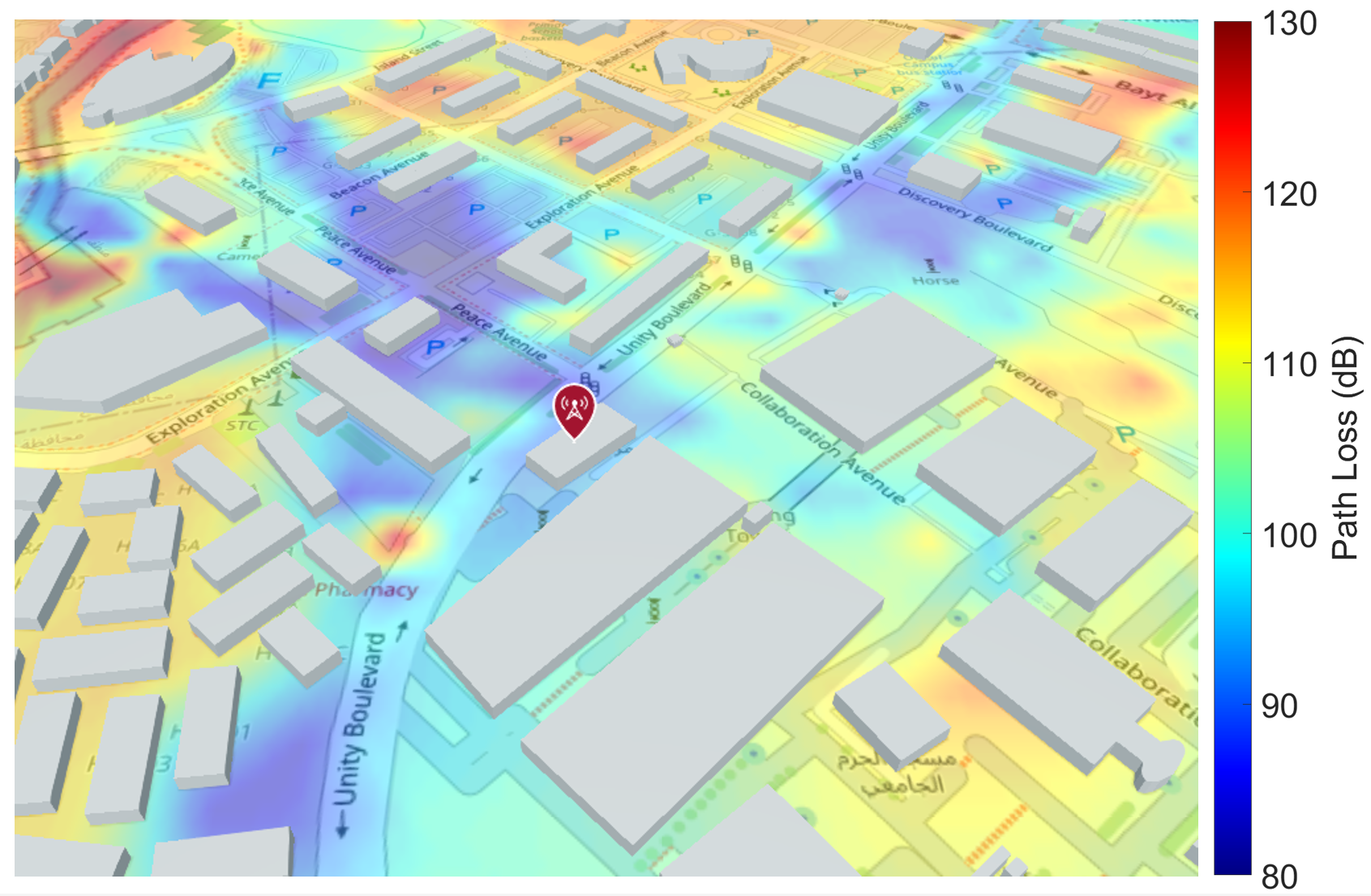}
    \\(b)
\end{minipage}
\caption{\textbf{\boldmath}The coverage maps of site A for $f=2.3\ \text{GHz}$, $P_{Tx}=5\ \text{W}$, $h_{Tx}=12\ \text{m}$: (a) signal strength in dBm and (b) path loss in dB.}
\label{maps1}
\end{figure}

\begin{figure}[ht]
\centering
\begin{minipage}{0.45\textwidth}
    \centering
    \includegraphics[width=85mm]{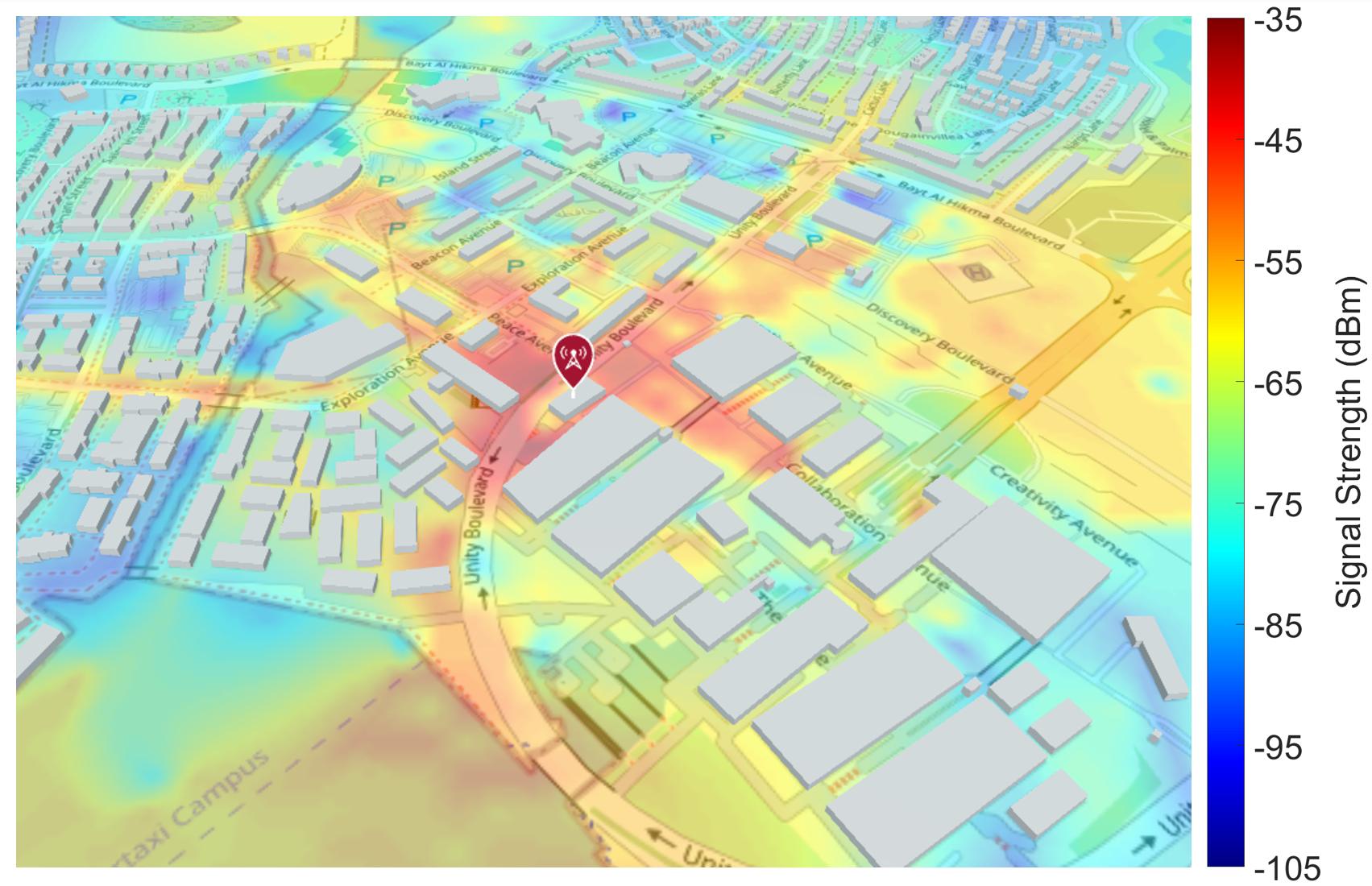}
    \\(a)
\end{minipage}
\hfill
\begin{minipage}{0.45\textwidth}
    \centering
    \includegraphics[width=85mm]{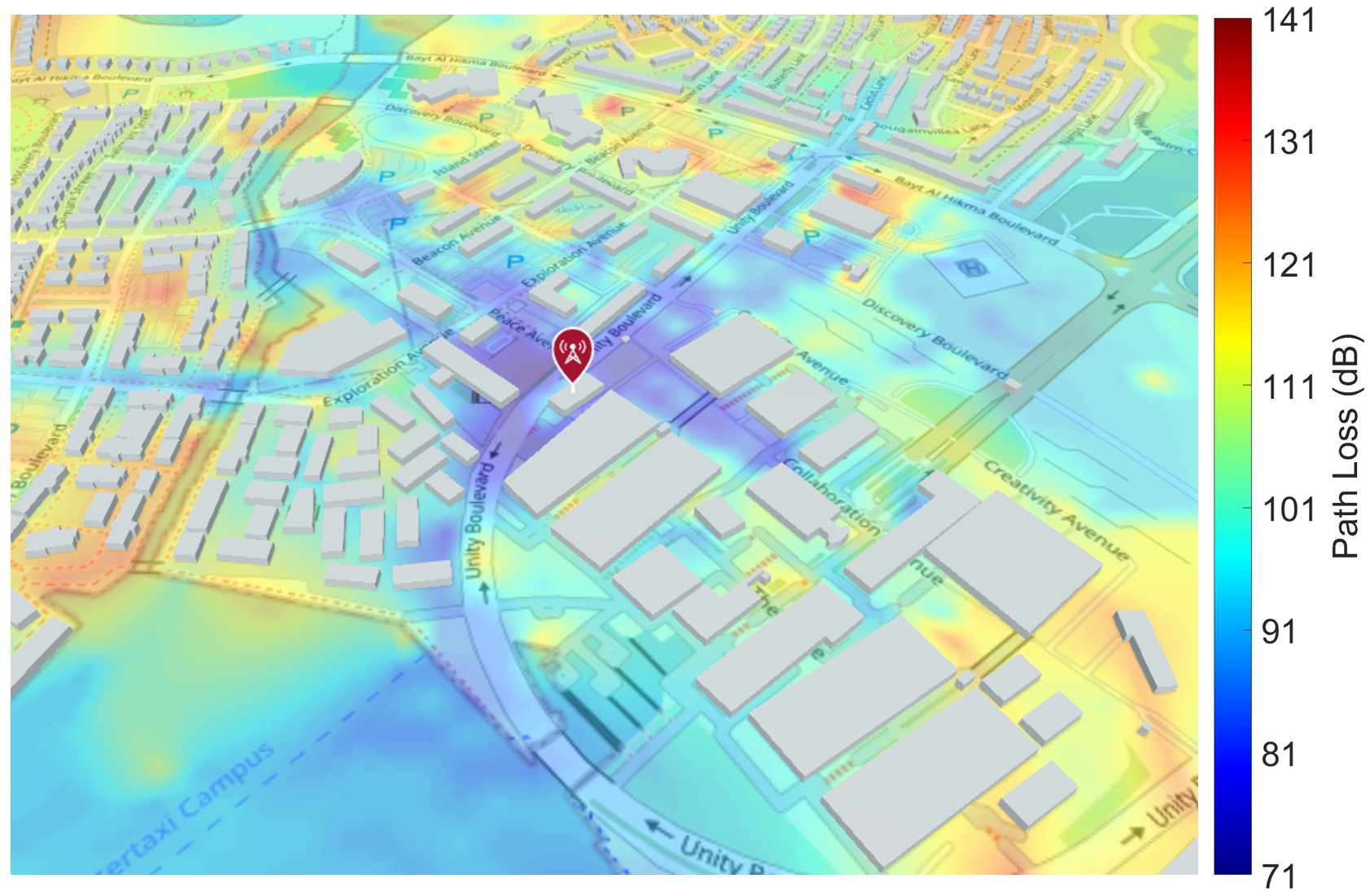}
    \\(b)
\end{minipage}
\caption{\textbf{\boldmath} The coverage maps of site A for $f=2.3\ \text{GHz}$, $P_{Tx}=5\ \text{W}$, $h_{Tx}=21\ \text{m}$: (a) signal strength in dBm and (b) path loss in dB.}
\label{maps10}
\end{figure}

For instance, Figure \ref{maps1} shows the obtained signal strength and PL in one sub-dataset of site A for the following parameters: $f=2.3$ GHz, $P_{Tx}=5$ W, $h_{Tx}=12 $ m, i.e., the height of the antenna above ground level is equal to the height of the mast (1 m) plus the height of the building (11 m). As expected, the received power decreases and consequently the attenuation increases as the distance between the transmitter and receiver increases. We notice that the coverage area is not necessarily circular, in other words, there are locations where ray-tracing did not compute the received signal. This behavior stems from the ray-tracing tool’s reliance on user-defined limits for reflections and diffractions, when no valid ray path can be found within those constraints, the model produces no power values for those locations. Moreover, placing the transmitter antenna on a taller tower leads to a wider coverage area and a higher received signal as shown in Figure \ref{maps10}. 

\vspace{-3mm}
\section{Results And Discussions}
\label{4}

\vspace{2mm}

\subsection{Models Setup}
In our ANN model, we use
4 hidden layers consisting of feed-forward neural networks. Each hidden layer is followed
by a sigmoid activation function. The first two hidden layers comprise 128 nodes, the third layer has 64 neurons, and the final layer consists of 32 neurons. We use a linear activation function with the output layer. Using the Optuna technique \cite{opt} with some trials, we set the learning rate as 0.001, and we use the Adam optimizer and the mean squared error (MSE) as loss function. For the RNN-LSTM model, we use two LSTM hidden layers with 400 and 50 nodes respectively, and one feed-forward layer with 200 units. We utilize the ReLU activation function, the Adam optimizer, the learning rate of 0.001, and the MSE as a loss function. We run these models using 1000 epochs. Concerning the DTR model, using Optuna we set the optimal value for the maximum tree depth as 30. We use 100 estimators and the maximum tree depth for the RFR model is set to 30. Finally, for the KNN, after evaluating various distance metrics and neighbor numbers 'K', we determined that using the Manhattan distance with K=30 delivers the best performance. To ensure the model's stability, we set the random seed to 42 to improve prediction reproducibility. Furthermore, we use the k-folds cross-validation method with k=10 in order to handle the overfitting problem. 

To evaluate the performance of each model, we measured the statistical error between the simulated and the predicted PL values using some well-known metrics including the root mean squared error (RMSE), the mean absolute percentage error (MAPE), the mean squared logarithmic error (MSLE), and the cross-correlation coefficient ($\rho$).
\begin{equation}\label{rmse}
 RMSE=\sqrt{\frac{1}{n} \sum\limits_{i=1}^n (y_i-\hat{y}_i)^2},
\end{equation}
\begin{equation}\label{mape}
 MAPE=\frac{100\%}{n} \sum\limits_{i=1}^n |\frac{y_i-\hat{y}_i}{y_i}|,
\end{equation}
\begin{equation}\label{msle}
 MSLE=\frac{1}{n} \sum\limits_{i=1}^n (\log(y_i+1)-\log(\hat{y}_i+1))^2,
\end{equation}
\begin{equation}\label{rho}
\rho=\frac{\sum\limits_{i=1}^n (y_i-\bar{y})(\hat{y_i}-\bar{\hat{y}})}{\sqrt{\sum\limits_{i=1}^n (y_i-\bar{y})^2 \sum\limits_{i=1}^n (\hat{y_i}-\bar{\hat{y}})^2}},
\end{equation}
where $y_i$ and $\hat{y_i}$ indicate respectively the actual and the predicted values for the $i^{th}$ data sample, n is the total number of samples, $\bar{y}$ is the mean value of $y$ and $\bar{\hat{y}}$ is the mean value of $\hat{y}$.

\vspace{-3mm}

\subsection{Results Using Simulated Test Dataset}
The error metrics are calculated by comparing the predicted PL using the trained  ML models, Longley-Rice model, and Close-in model with the simulated test dataset as listed in Table \ref{errors}. As we observe, the values of the statistical errors of the examined ML approaches are notably low compared to the empirical models. The results show that the ML models yield satisfactory results with RMSE values ranging between 1.35 dB and 4.62 dB which is within the acceptable range in the order of 6-7 dB \cite{15}. We notice that the models have better performance in the prediction of the LoS PL compared to NLoS scenarios. This may be explained by the complex propagation phenomena in the case of the NLoS scenarios which result in increased variability.
Furthermore, our dataset contains a higher number of LoS samples compared to NLoS samples. This difference in sample distribution can lead to higher accuracy in the LoS situations. 

\begin{figure}[ht]
\centering
\includegraphics[width=90mm]{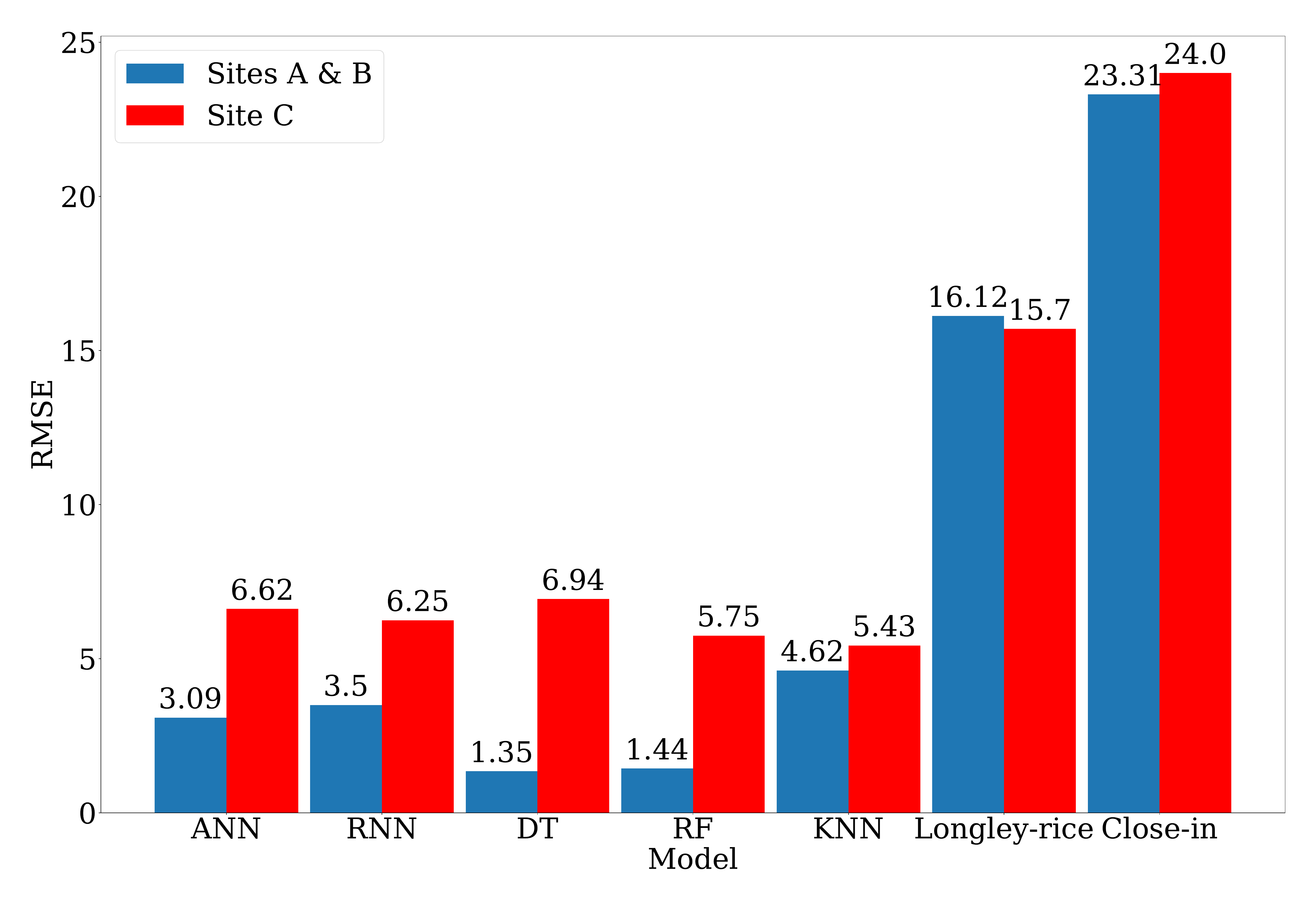}
    \vspace{-1em}
    \caption{The RMSE values for the examined models in the sites A, B, and C.}
    \label{fig:rmse}
\end{figure}

\begin{figure}[ht]
\centering
\includegraphics[width=90mm]{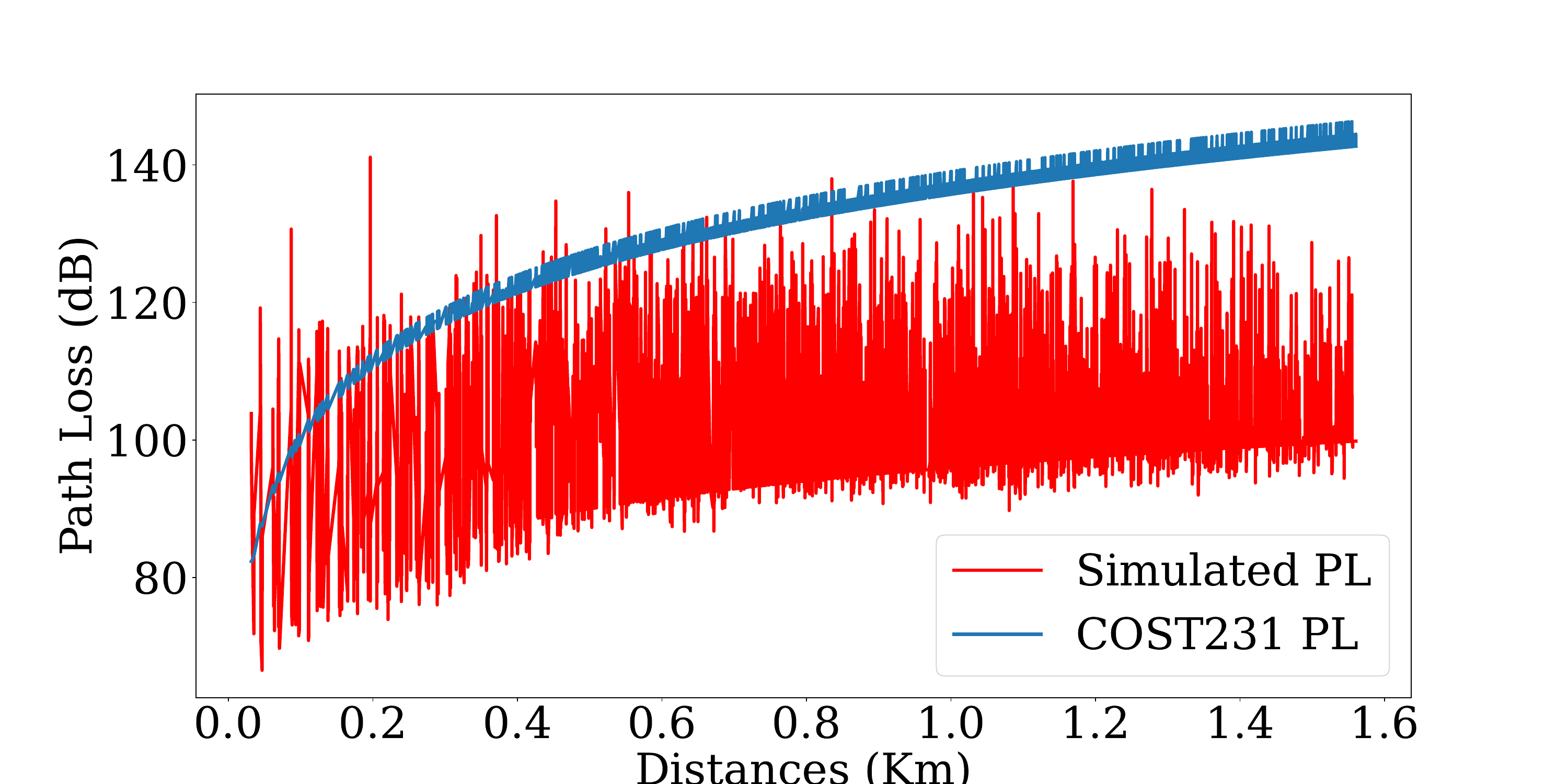}
\vspace{-1em}
\caption{The path loss as a function of distance using the COST231-Hata model for sites A and B for $f=1.5 \ \text{GHz}$.}
\label{AB}
\end{figure}

According to the results, DTR and RFR algorithms outperform the other ML models in LoS and NLoS scenarios. Although the KNN model accurately predicts the PL in the LoS case, it demonstrates the lowest performance in the NLoS scenario. ANN and RNN-LSTM show comparable results with ANN being slightly better. On the other hand, the statistical errors show that the Longley-Rice and Close-in models have the highest RMSE values of 16.12 dB and 23.31 dB respectively.
\begin{table*}[ht]
\scriptsize
\caption{Performance evaluation of the ML models and conventional PL models on 20\% of the datasets of Site A and Site B.}\label{errors}
\begin{tabular}{|p{0.6cm}|p{0.6cm}|p{0.6cm}|p{0.6cm}|p{0.6cm}|p{0.6cm}|p{0.6cm}|p{1.5cm}|p{1.5cm}|p{1.5cm}|p{0.9cm}|p{0.9cm}|p{0.9cm}|p{0.6cm}|p{0.6cm}|p{0.6cm}|p{0.6cm}|} 
\hline
~~~~Model &\multicolumn{3}{c|}{RMSE} &\multicolumn{3}{c|}{MAPE (\%)} &\multicolumn{3}{c|}{MSLE} &\multicolumn{3}{c|}{$\rho$} \\ 
\cline{2-13}
                    &LoS     &NLoS  &Total                   &LoS     &NLoS  &Total          
                    &LoS     &NLoS  &Total          
                    &LoS     &NLoS  &Total 
                    \\               
\hline
\multicolumn{1}{|c|}{ANN}&1.73  &4.79 &3.09  &0.84 &2.90 &1.52  &$9.5\times10^{-6}$ &$5.5\times10^{-5}$ &$2.4\times10^{-5}$ &0.961 &0.883  &0.953 \\ 
\hline
\multicolumn{1}{|c|}{RNN-LSTM}&1.77  &5.55 &3.50  &0.88 &3.53 &1.75 &$9.9\times10^{-6}$ &$7.3\times10^{-5}$ &$3.1\times10^{-5}$ &0.959 &0.840 &0.939 \\ 
\hline

\multicolumn{1}{|c|}{\textbf{DTR}} &\textbf{0.55}  &\textbf{2.22} &\textbf{1.35}  &\textbf{0.06} &\textbf{0.35} &\textbf{0.15}  &$\boldsymbol{9.7 \times 10^{-7}}$& $\boldsymbol{1.1 \times 10^{-5}}$ &$\boldsymbol{4 \times 10^{-6}}$ &\textbf{0.996} &\textbf{0.976} &\textbf{0.991}                 \\ 
\hline
\multicolumn{1}{|c|}{RFR} &0.66  &2.33 &1.44 &0.24 &1.12 &0.53  &$1.4\times10^{-6}$ &$1.3\times10^{-5}$ &$5\times10^{-6}$ &0.994 &0.975  &0.990      \\ 
\hline
\multicolumn{1}{|c|}{KNN} &1.90  &7.58 &4.62 
 &0.84 &5.08 &2.24  &$1.1\times10^{-5}$ &$1.3\times10^{-4}$ &$5.2\times10^{-5}$ &0.953 &0.670   &0.891       \\ 
\hline
\multicolumn{1}{|c|}{Longley-Rice} &5.34  &27.02 &16.12  &4.16 &18.79 &8.98  &$8.5\times10^{-5}$ &$1.2\times10^{-3}$ &$4.6\times10^{-4}$  &0.817 &0.251  &0.683 \\ 
\hline
\multicolumn{1}{|c|}{Close-in} &27.37  &11.21 &23.31 &26.81 &8.52 &20.78  &$1.7\times10^{-3}$ &$2.7\times10^{-4}$ &$1.2\times10^{-3}$  &0.933 &0.568  &0.301                \\ 
\hline
\end{tabular}
\end{table*}
We evaluate the same models using new data collected from site C. As shown in Figure \ref{fig:rmse}, the ML algorithms can still provide acceptable results at different suburban locations i.e., RMSE $\leq$ 7dB, widely used as the benchmark threshold for path loss prediction in wireless communication systems \cite{15}, indicating satisfactory performance as mentioned earlier with the KNN model being the most accurate in site C. This can be attributed to the nature of the KNN algorithm, which is an instance-based method that predicts the received signal strength at a new point by averaging the values of the 'K' most similar training samples, as explained previously in its description. It performs well in capturing localized spatial patterns, as receivers in close proximity or at comparable distances from the BS under similar conditions, often exhibit similar signal strengths. Unlike parametric models, KNN relies on memorizing training data and interpolating between points, which makes it effective for complex and nonlinear propagation environments. However, the conventional models exhibit a limited ability to predict the signal attenuation with a large RMSE value of 15.703 dB. 
Furthermore, we evaluate the prediction accuracy of the COST231-Hata model for sites A and B. We evaluate this model at 1.5 GHz which satisfies the model constraints. Figure \ref{AB} shows the predicted PL as a function of distance using both the COST231-Hata model and the simulated data. As we can notice, the examined COST231-Hata model generally tends to over-predict the PL. This observation is further supported by the
significant RMSE value of 35.579 dB. However, we have to mention that we deviate from some restrictions by using a transmitter height of approximately 20 m, whereas the minimum height specified is 30 m for this model. In addition, the model is applicable for distances larger than 1 km while we calculate the PL for a distance up to 1.5 km. Although these results are not reported on the overall data, they indicate significant errors compared to those obtained by ML models. Furthermore, we test the COST231-Hata model using these configurations for site C at f=1.5 GHz which results in a RMSE of 36.456 dB. We conclude that this model is unable to accurately predict PL. Indeed, this analysis reinforces the effectiveness of ML-based models in PL prediction.

\vspace{-3mm}

\subsection{Results Using Measured Test Dataset}
In this section, the performance of ML models in the prediction of the PL is evaluated compared to measured data given in \cite{13}. The measurement campaign was carried out in Rio de Janeiro, Brazil and used to evaluate the performance of a hybrid propagation model for macrocells. The measured data represent the received signal strength corresponding to two distinct receiver routes with different transmitter locations, at frequencies 750 MHz, 2.5 GHz and 3.5 GHz. The parameters used in the measurement campaign at different frequencies are listed in Table \ref{systparamet}. 
\begin{table}[ht]
\centering
\caption{The measurement campaign parameters \cite{13}.}
\begin{tabular}{|l|c|c|c|}
\hline
\textbf{Parameter}     & \textbf{750 MHz} & \textbf{2.5 GHz} & \textbf{3.5 GHz}
\\\hline
Transmitter power (dBm)            &10  &10  &10\\\hline
Transmitter antenna gain (dBi)     &5 &5  &5\\ \hline
Receiver antenna gain (dBi) &1   &5  &6\\ \hline
\end{tabular}
\label{systparamet}
\end{table}

Two distinct scenarios were adopted to collect this data. The geographical coordinates of the transmitter antenna (Tx) in each case are illustrated in Table \ref{txloc}. In the first scenario SC1, the transmitter antenna of height 2 m was installed on the top of a building and covers a distance of approximately 22.4 km while the receiver is located at a height of 3 m. A total of 3,887 samples were obtained for each frequency value. On the other hand, in the second scenario SC2, the transmitter was placed on the top of a hill, standing at a height of 3 m. The receiver was located along the coastal road and the data were collected up to 6 km. Around 1,459 samples were provided for each frequency in this scenario. The maximum separation distance recorded in SC1 was 3.45 km, while it was 1.9 km for SC2.

\begin{figure}[ht]
\centering
\includegraphics[width=85mm]{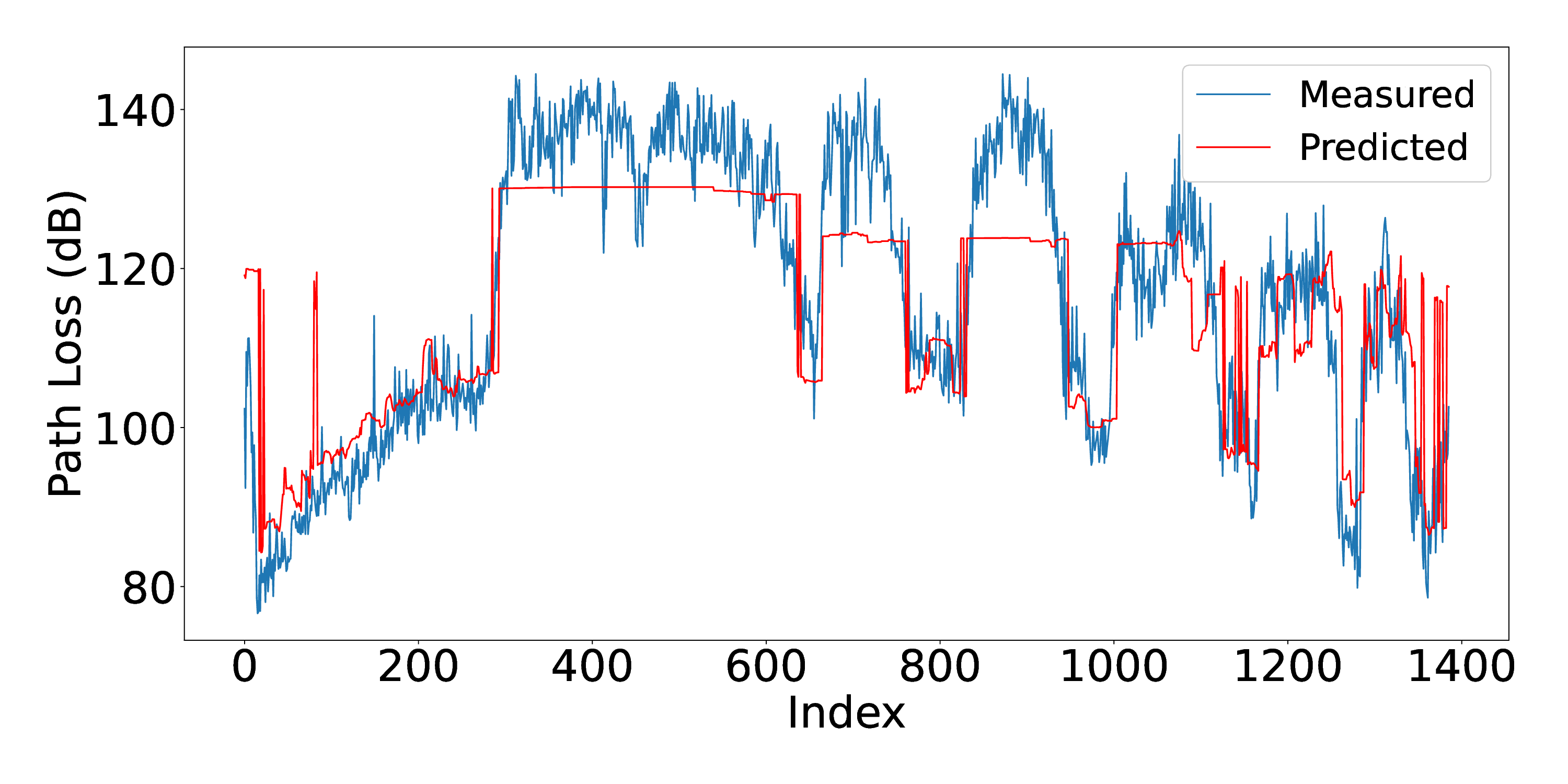}
\vspace{-1em}
\caption{Sample of the predicted and measured path loss data for the second scenario (SC2) at a frequency of 3500 MHz.}
\label{fig:sc2}
\end{figure}

\begin{table}[ht]
\centering
\caption{Transmitter Locations.}
\begin{tabular}{|p{0.28\linewidth}| p{0.28\linewidth}| p{0.28\linewidth}|}
\hline
\textbf{Scenario}     & \textbf{Tx latitude} & \textbf{Tx longitude} 
\\\hline
SC1            & -22.9795°  & -43.2319°  \\\hline
SC2    & -22.9892°  & -43.2290° \\ \hline
\end{tabular}
\label{txloc}
\end{table}
\begin{table*}[ht!]
\centering
\caption{Performance evaluation of the RFR model for all scenarios at different frequencies.}
\label{measerror}
\begin{tabular}{ |l|l|l|l|l|l|c| }
\hline
Scenario & Frequency (MHz) & RMSE &MAPE (\%) &MSLE &$\rho$\\ \hline
\multirow{3}{*}{SC1} & 750  & 11.01 & 8.41 & $3.37\times10^{-4}$ & 0.754 \\
 & 2500  & 9.73 & 6.29 & $2.05\times10^{-4}$ & 0.783 \\
 & 3500  & 11.32 & 7.64 & $2.78\times10^{-4}$ & 0.699 \\ \hline
\multirow{3}{*}{SC2} & 750  & 9.80 & 7.81 & $3.14\times10^{-4}$ & 0.864 \\
 &2500  & 9.58 & 6.46 & $2.02\times10^{-4}$ & 0.892 \\
 &3500 & 9.27 & 6.30 & $2.08\times10^{-4}$ & 0.861 \\ \hline
\end{tabular}
\end{table*}
We used the available data in \cite{leonor2022site} including the antenna coordinates, heights, and separation distances with the available information in \cite{13} to extract the required features to test the ML models. For comparison purposes, we use the experimental PL values obtained during the measurements, available in \cite{leonor2022site}.
In Table~\ref{measerror}, we evaluated the different ML models based on the measured data and we summarize the best results for the two scenarios. The reported errors are for the RFR model as it outperforms the other models for all cases. This can be explained by the fact that the signal propagation is highly complex and nonlinear, influenced by factors such as obstacles, buildings, vegetation, and terrain topography. As a result, PL cannot be accurately modeled as a simple function of distance, and the real-world dataset reflects complex nonlinear relationships between input features and the PL. RFR, as an ensemble of decision trees which mitigates overfitting, as mentioned previously, naturally handles such nonlinearity without requiring feature transformations and generalizes well on unseen samples, making it well suited for this scenario. Despite having relatively high RMSE errors ranging from 9.27 dB to 11.32 dB, exceeding 7 dB, our results remain acceptable, particularly when compared to other models discussed in \cite{13}. For instance, Okumura Hata, log-distance, and Longley-Rice yield mean RMSE values of 17.25 dB, 13.63 dB, and 38.45 dB respectively as indicated in \cite{13}. Validation using the same measurement dataset in \cite{13} shows that our ML-based model outperforms the proposed model in \cite{13}, achieving a lower RMSE of 10.11 dB compared to their reported range of 11.50–13.46 dB. The MAPE and MSLE metrics exhibit values within the range of  6.29-8.41\% and $2,02.10^{-4}-3,37.10^{-4}$. 
Overall, the RFR model reveals a satisfactory ability to predict the PL with slightly better results in SC2 according to the performance metrics. However, referring to Figure \ref{fig:sc2}, it is worth mentioning that in some locations in SC2, the model tends to underestimate the PL. 

In that regard, we can resort to the explanation provided in \cite{leonor2022site} which indicates that in SC2, the receiver experiences a rapid transition from LoS to NLoS position, where the direct path between the two antennas is obstructed by multiple buildings. This situation arises because the transmitter was not elevated significantly over the rooftops of the nearby structures. These results are expected since we have trained the ML models on simulated data without taking into account the impact of the exact building structures and terrain profiles as well as the dense vegetation. Figure \ref{fig:sc2} illustrates the most accurate results achieved in SC2 for 3500 MHz. The results indicate that our proposed ML model (RFR), trained on simulations, outperforms other empirical and hybrid models. Consequently, the ray-tracing tool is an effective option for generating synthetic datasets to train ML models due to its flexibility and adaptability to real-world scenarios.

\vspace{-3mm}

\section{Conclusion}
\label{5}
In this work, we examined the performance of several ML models including ANN, RNN-LSTM, DTR, RFR, and KNN in predicting the PL in different suburban sites on the KAUST campus for the sub-6 GHz frequency band. We utilized the ray-tracing simulation technique to generate the training and testing datasets for different conditions and configurations. We compared the ML-based approaches to traditional propagation models including Longley-Rice, close-in, and COST231-Hata models. Our results demonstrated the superiority of ML-based propagation models over conventional models. Notably, the DTR model exhibited the best performance in sites A and B, while KNN  outperformed the other models in site C. We also verified and validated the developed ML-based models with the measurement data given in \cite{13}. 
The results show that the RFR-based model, trained using synthetic datasets, achieved better performance compared to the proposed models in \cite{leonor2022site}. Overall, the results showed that the ML algorithms are a promising solution for predicting PL, as they achieved satisfactory performance compared to simulated and measurement test data. Additionally, utilizing the ray-tracing tool to build ML-based PL models is an attractive, flexible, and cost-effective option. It should be noted that training ML models, particularly DL models such as ANN and RNN, can be time-consuming due to hyperparameter tuning and performance optimization. Nevertheless, once trained, predictions on new samples can be generated in seconds using the trained weights, allowing near-instant inference on standard consumer-grade computers.

In this work, we did not consider dynamic environments and vegetation loss. We also used a simplified representation of the exterior structure of buildings and houses. These factors can affect the accuracy of PL prediction. We will incorporate vegetation loss in future works, as it plays a critical role in the accuracy of PL prediction, especially in areas with dense vegetation. We also aim to take into account the dynamic nature of the environment and include more realistic terrain profiles and building structures. To this end, employing a more advanced version of the ray-tracing tool represents a promising future direction, as it could improve accuracy by incorporating higher-order diffractions and reflections and accounting for real-world environmental dynamics. However, increased computational complexity should be expected in more intricate scenarios. Consequently, a high-performance computer equipped with a graphics processing unit (GPU) would be required to run the ray-tracing tool for dataset generation. Additionally, extending the training of ML models to a wider range of samples collected under diverse conditions could enhance the generalizability and flexibility of the ML-based models.

\section*{Acknowledgments}
 This work was supported by the ERIF/OSSARI Funding.

\bibliographystyle{IEEEtran}
\bibliography{sample}

\vspace{-10mm}

\begin{IEEEbiography}
[{\includegraphics[width=1in,height=1.25in,clip,keepaspectratio]{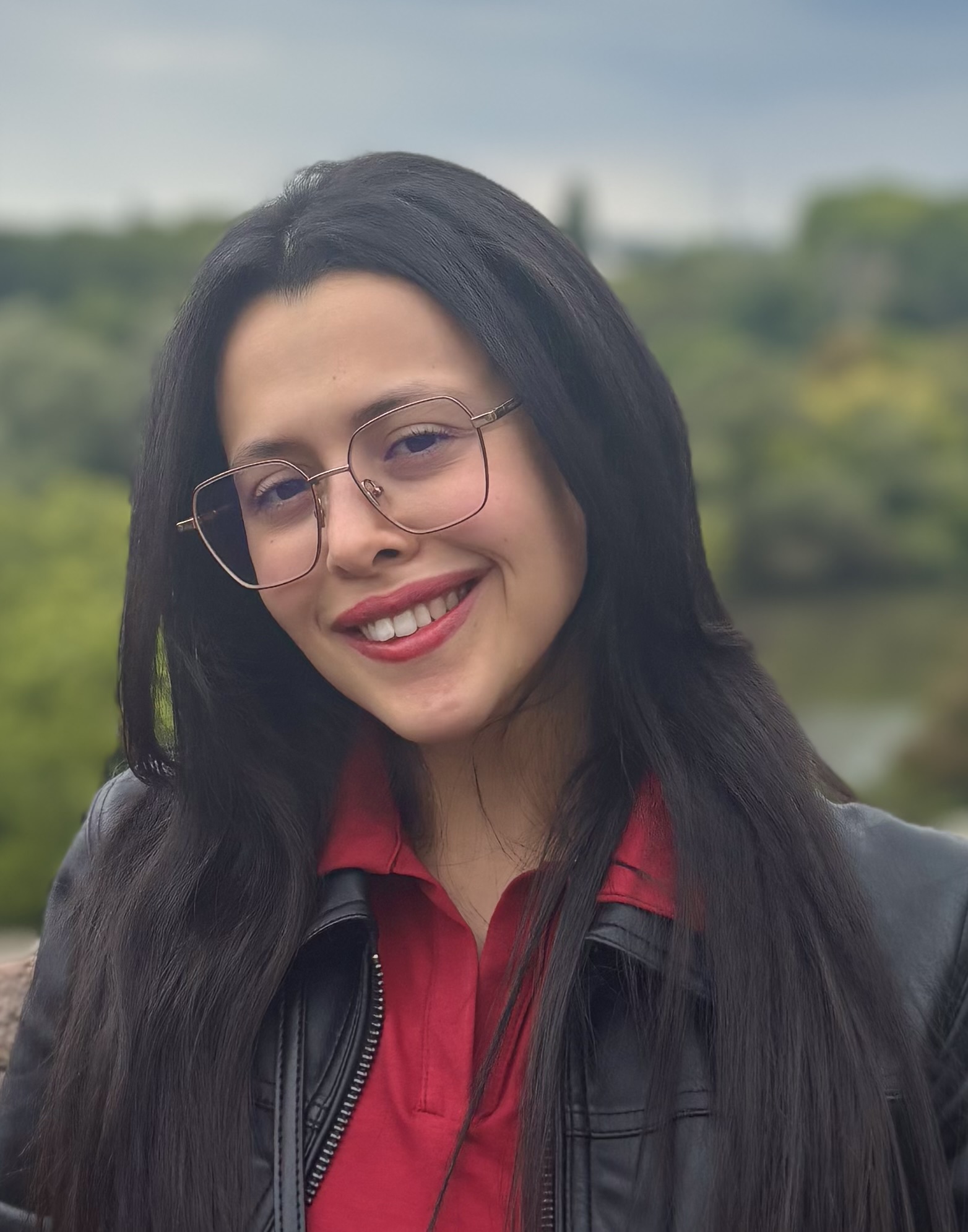}}]
{FERDAOUS TARHOUNI}~{}received the National Engineering Diploma degree from the Ecole Polytechnique de Tunisie (EPT), and the M.Sc. degree in electrical and computer engineering from the King Abdullah University of Science and Technology (KAUST), Thuwal, Saudi Arabia. She is currently pursuing the Ph.D. degree with the Institute for Digital Communications (IDC), Friedrich-Alexander-Universit\"at Erlangen-N\"urnberg, Germany. Her current research interests include non terrestrial networks, free space optical communication, stochastic geometry, and reconfigurable intelligent surfaces.
\end{IEEEbiography}

\vspace{-10mm}

\begin{IEEEbiography}[{\includegraphics[width=1in,height=1.25in,clip,keepaspectratio]{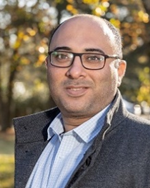}}]
{MUNEER AL-ZUBI}~{}~received the Ph.D. degree in engineering from the University of Technology Sydney (UTS), Sydney, Australia, in 2020. He worked as a Research Associate with the Center of Excellence for Innovative Projects, Jordan University of Science and Technology (JUST), from 2020 to 2021. From 2021 to 2022, he was a postdoctoral researcher with the Department of Engineering, University of Luxembourg, Luxembourg, and also, he was a remote visiting scholar with the School of Electrical and Data Engineering, UTS. He is currently a Postdoctoral Researcher at the Communication Theory Lab (CTL), King Abdullah University of Science and Technology, Saudi Arabia. His research interests lie in the areas of wireless communication, EM wave propagation, and molecular communication.
\end{IEEEbiography}

\vspace{-10mm}

\begin{IEEEbiography}[{\includegraphics[width=1in,height=1.25in,clip,keepaspectratio]{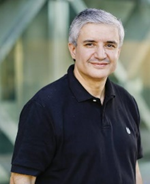}}]
{MOHAMED-SLIM ALOUINI}~{(S’94, M’98, SM’03, F’09)}~received the Ph.D. degree in electrical engineering from the California Institute of Technology (Caltech), Pasadena, CA, USA, in 1998. He served as a faculty member at the University of Minnesota, Minneapolis, MN, USA, then at Texas A\&M University at Qatar, Education City, Doha, Qatar before joining King Abdullah University of Science and Technology (KAUST), Thuwal, Makkah Province, Saudi Arabia as a professor of electrical engineering in 2009. His current research interests include the modeling, design, and performance analysis of wireless communication systems.
\end{IEEEbiography}

\vfill\pagebreak

\end{document}